\documentclass{ptptex}

\usepackage{graphicx}

\title{%
Transient behavior of a population dynamical model

}

\author{%
B. Spagnolo \footnote{E-mail: spagnolo@unipa.it}, D. Valenti, A.
Fiasconaro}

\inst{%
Dipartimento di Fisica e Tecnologie Relative and INFM,\\ Group of
Interdisciplinary Physics, Universita' di Palermo \\ Viale delle
Scienze, I-90128 Palermo, Italy}

\abst{%

The transient behavior of an ecosystem with N random interacting
species in the presence of a multiplicative noise is analyzed. The
multiplicative noise mimics the interaction with the environment.
We investigate different asymptotic dynamical regimes and the role
of the external noise on the probability distribution of the local
field.}

\begin{document}

\maketitle

\section{Introduction}
Population dynamics attracted a lot of attention in recent years
and became the object of many studies as well by biologists as by
physicists \cite{Mur93,Kaw02,Gia01,Esc04,Droz04}. Tools developed
in the context of nonequilibrium statistical physics to analyze
nonequilibrium nonlinear physical systems provide new insights and
at the same time new approaches to study biological systems.
Biological population dynamics has many interesting, and still not
solved, problems such as pattern formation
\cite{Soc01,Bar01,Val04,Fia04}, the role of the noise on complex
ecosystem behaviour, and the noise-induced effects, such as
stochastic resonance, noise delayed extinction, quasi periodic
oscillations etc...
\cite{Vil98,Spa02,Roz01,Spa03,Valb04,Sci99,Sci01,Spa04,Ras00}. The
dynamical behavior of ecological systems of interacting species
evolves towards the equilibrium states through the long, slow, and
complex process of nonlinear relaxation, which is strongly
dependent on the random interaction between the species, the
initial conditions and the random interaction with environment. A
good mathematical model to analyze the dynamics of $N$ biological
species with spatially homogeneous densities is the generalized
Lotka-Volterra system with a Malthus-Verhulst modelization of the
self regulation mechanism, in the presence of a multiplicative
noise \cite{Ciu96,Spab02,Cir03}. By neglecting the fluctuations of
the local field we derive a quasi-stationary probability of the
populations. We obtain the asymptotic analytical expressions for
different nonlinear relaxation regimes, and we analyze the role of
the multiplicative noise on the probability distribution of the
local field.

\section{The model and results}

The dynamical evolution of our ecosystem composed by $N$
interacting species in a noisy environment and in the presence of
an absorbing barrier is described by the following Ito stochastic
differential equation

\begin{equation}
d x_i(t) = \left[ \left(g_i(x_i(t)) + \sum_{j\neq i} J_{ij}x_j(t)
\right)dt + \sqrt{\epsilon} dw_i\right] x_i(t)\mbox{,} \enspace
\thinspace i = 1,...,N \label{langevin}
\end{equation}
where $x_i(t) \geq 0$ is the population density of the
\textit{$i^{th}$} species at time $t$ and the function
$g_i(x_i(t))$

\begin{equation}
g_i(x_i(t)) = \left(\alpha + \frac{\epsilon}{2} \right) - x_i(t)
\label{g function}
\end{equation}
describes the development of the \textit{$i^{th}$} species without
interacting with other species. In Equation (\ref{langevin}),
$\alpha$ is the growth parameter, the interaction matrix $J_{ij}$
models the interaction between different species ($i\neq j$), and
$w_i$ is the Wiener process whose increment $dw_i$ satisfies the
usual statistical properties $<dw_i(t)> \thinspace = \thinspace
0$, and $<dw_i(t)dw_j(t^{\prime})> \thinspace = \thinspace
\delta_{ij}\delta(t-t^{\prime}) dt$. We consider a random
asymmetric interaction matrix $J_{ij}$, whose elements are
independently distributed according to a Gaussian distribution
with $\langle J_{ij}\rangle = 0$, $\langle J_{ij} J_{ji}\rangle =
0$, and $\sigma^2_j = J^2/N$. The term $J_{ij}x_i x_j$ is the loss
or the growth rate of species $i$ due to interaction with species
$j$, when $J_{ij} < 0$ or $J_{ij} > 0$ respectively. With this
choice of interaction matrix our ecosystem contains 50$\%$ of
prey-predator interactions ($J_{ij} < 0$ and $J_{ij} > 0$), 25$\%$
competitive interactions ($J_{ij} < 0$ and $J_{ij} < 0$), and
25$\%$ symbiotic interactions ($J_{ij} < 0$ and $J_{ij} > 0$). We
consider all species equivalent so that the characteristic
parameters of the ecosystem are independent of the species. The
random interaction with the environment (climate, disease,etc...)
is taken into account by introducing a multiplicative noise in the
Equation (\ref{langevin}). The solution of the dynamical Equation
(\ref{langevin}) is given by

\begin{equation} x_i(t) = \frac{x_i(0) z_i (t)} {1+
x_i(0) \int_{0}^{t}dt^{\prime}  z_i (t^{\prime})},
\label{sol-langevin}
\end{equation}
where

\begin{equation} z_i(t) = e^{\alpha t +\sqrt{\epsilon} w_i(t) +
\int_{0}^{t} dt^{\prime}\eta_i(t^{\prime})} \label{z-process}
\end{equation}
and

\begin{equation}
\eta_i(t) = \sum_{j\neq i}J_{ij}x_j(t) \label{local field}
\end{equation}
is the local field acting on the \textit{$i^{th}$} population and
represents the influence of other species on the differential
growth rate. We note that the dynamical behavior of the
\textit{$i^{th}$} population depends on the time integral of the
process $z_i(t)$ and the time integral of the local field. For a
large number of interacting species we can assume that the local
field $\eta_i(t)$ is Gaussian with zero mean and variance
$\sigma_{\eta_i}^2 = \langle \eta_i^2\rangle = J^2 \langle x_i^2
\rangle$. As a consequence, in the absence of external
multiplicative noise, from the fixed-point equation $x_i(\alpha -
x_i + \eta_i) = 0$, the stationary probability distribution of the
populations is the sum of a truncated Gaussian distribution at
$x_i =0$ ($x_i >0$ always) and a delta function for extinct
species. The initial values of the populations $x_i(0)$ have also
Gaussian distribution with mean value $\langle x_i(0)\rangle = 1$,
and variance $\sigma^2_{x(0)} = 0.01,0.03,0.05$.

The interaction strength between the species $J$ determines two
different dynamical behaviors of the ecosystem. Above a critical
value $J_c = 1.1$, the system is unstable and at least one of the
populations diverges. Below $J_c$ the system is stable and
asymptotically reaches an equilibrium state. The equilibrium
values of the populations depend both on their initial values and
on the interaction matrix. If we consider a quenched random
interaction matrix, the ecosystem has a great number of
equilibrium configurations, each one with its attraction basin.
For an interaction strength $J = 1$ and an intrinsic growth
parameter $\alpha = 1$ we obtain: $<x_i> = 1.4387, <x^{2}_i> =
4.514,$ and $\sigma^{2}_{x_i} = 2.44$. These values agree with
that obtained from numerical simulation of Equation
~(\ref{langevin}). From the Fokker-Planck equation associated to
the Langevin equation (\ref{langevin})

\begin{equation}
\frac{\partial}{\partial t}P(x_i,t) = - \frac{\partial}{\partial
x_i}\left[ \frac{\epsilon}{2}\frac{\partial}{\partial x_i} x_i^{2}
- (\alpha + \frac{\epsilon}{2} - x_i + \eta_i)x_i \right] P(x_i,t)
\label{FP-eq.}
\end{equation}
we obtain a quasi stationary distribution by neglecting the
fluctuations of the local field in the asymptotic regime

\begin{equation}
\frac{d P(x_i)}{P(x_i)} = \frac{2}{\epsilon}\left[\alpha + \eta_i
- x_i - \frac{\epsilon}{2}\right]\frac{d x_i}{x_i} \label{Pi}
\end{equation}
that is

\begin{equation}
 P(x_i) = N_{x_i(0)}exp\left[\frac{2}{\epsilon}\left((\alpha + \eta_i -
 \frac{\epsilon}{2})\ln{x_i} - x_i\right)\right]\Theta(x_i)
\label{Pi(ni)}
\end{equation}
where the normalization factor is
\begin{equation}
N_{x_i(0)} =  \frac{P_{x_i(0)} e^{\frac{2x_i(0)}{\epsilon}}}
 {x_i(0)^{\frac{2}{\epsilon}(\alpha + \eta_i -
 \frac{\epsilon}{2})}}
 \label{Ni(0)}
\end{equation}
and $\Theta$ is the Heaviside unit step function. Now we focus on
the statistical properties of the time integral of the {\em
$i^{th}$} population $X_i(t)$

\begin{equation} X_i(t) = \int_{0}^{t} dt^{\prime} x_i(t^{\prime}),
\label{Xi(t)}
\end{equation}
\\
in the asymptotic regime. From Eqs. (\ref{sol-langevin}) and
(\ref{Xi(t)})we have

\begin{equation} X_i(t) =
\ln\left[1+ x_i(0) \int_{0}^{t} dt^{\prime} exp\left[\alpha
t^{\prime}+\sqrt{\epsilon} w_i(t^{\prime}) +
\eta_i(t^{\prime})\right]\right] ,
\label{Xi Integr. Equation}
\end{equation}
\\
We  use the same approximation of the mean field interaction
\cite{Ciu96}, and after differentiating Eq.(\ref{Xi Integr.
Equation}), we get the asymptotic solution of $X_i(t)$ as

\begin{equation} X_i(t) \simeq
 \ln\left[ x_i(o) e^{\sqrt{\epsilon} w_{max_i}(t) +
\eta_{max_i}(t)} \int^t_0 dt^{\prime} e^{\alpha t^{\prime}}\right]
\label{Xi a>0 =0}
\end{equation}
\\
where $w_{max_i}(t) = sup_{0<t^{\prime}<t}w(t^{\prime})$ and $
\eta_{max_i}(t) = sup_{0<t^{\prime}<t}\eta(t^{\prime})$. Eq.
(\ref{Xi a>0 =0}) is valid for $\alpha \geq 0$, that is, when the
system relaxes toward an equilibrium population and at the
critical point. After making the ensemble average, we obtain for
the time average of the {\em $i^{th}$} population $\bar{X_i}$

\begin{equation}
\left<\bar{X_i}\right> \simeq \frac{1}{t} \left[N_w \sqrt{\epsilon
t} + \ln t + \left<\ln\left[n_i(o) \right]\right>\right] \mbox{,}
\,\,\, \alpha = 0, \label{Xi a=0}
\end{equation}
and
\begin{equation}
\left<\bar{X_i}\right> \simeq \frac{1}{t} \left[N_w \sqrt{\epsilon
t} + (\alpha + N_{\eta})t + \left<\ln\left[\frac{x_i(o)}{\alpha}
\right]\right>\right] \mbox{,} \,\,\, \alpha > 0, \label{Xi a>0}
\end{equation}
\\
where $N_w$ and $N_{\eta}$ are variables with a semi-Gaussian
distribution~\cite{Ciu96} and $N_{\eta}$ must be determined
self-consistently from the variance of the local field (Eq.
(\ref{local field})). We obtain, consistently with mean field
approximation, the typical long time tail behavior ($t^{-1/2}$)
dependence, which characterizes nonlinear relaxation regimes when
$\alpha \geq 0$. When the system relaxes toward the absorbing
barrier ($\alpha<0$), the time average of the {\em $i^{th}$}
population $\left<\bar{X_i}\right>$ is a functional of the local
field and the Wiener process. We have also analyzed the dynamics
of the ecosystem when one species is absent. Specifically, we
considered the cavity field, which is the field acting on the {\em
$i^{th}$} population when this population is absent. The
probability distributions for both local and cavity field have
been obtained by simulations for a time $t = 200$ (a. u.) in
absence of external noise, and for different species. We found
that the probability distributions of the cavity fields differ
substantially from those of local fields for the same species,
while in the presence of noise the two fields overlap (see Figs. 1
and 2).
\begin{figure}[htb]
\parbox{\halftext}{
\includegraphics[width=7 cm,height=3 cm]{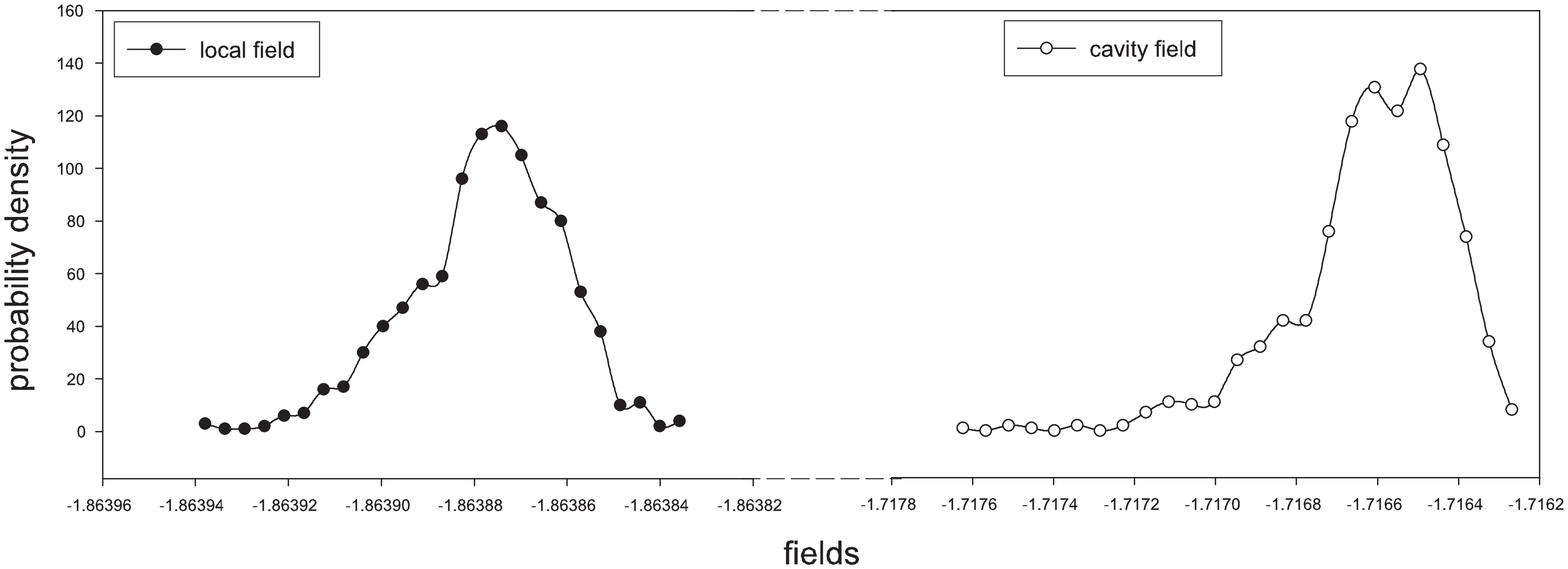}
\caption{The local and the cavity fields in the absence of noise,
$\epsilon = 0$.}} \hfill
\parbox{\halftext}{
\includegraphics[width=6.8 cm,height=2.8cm]{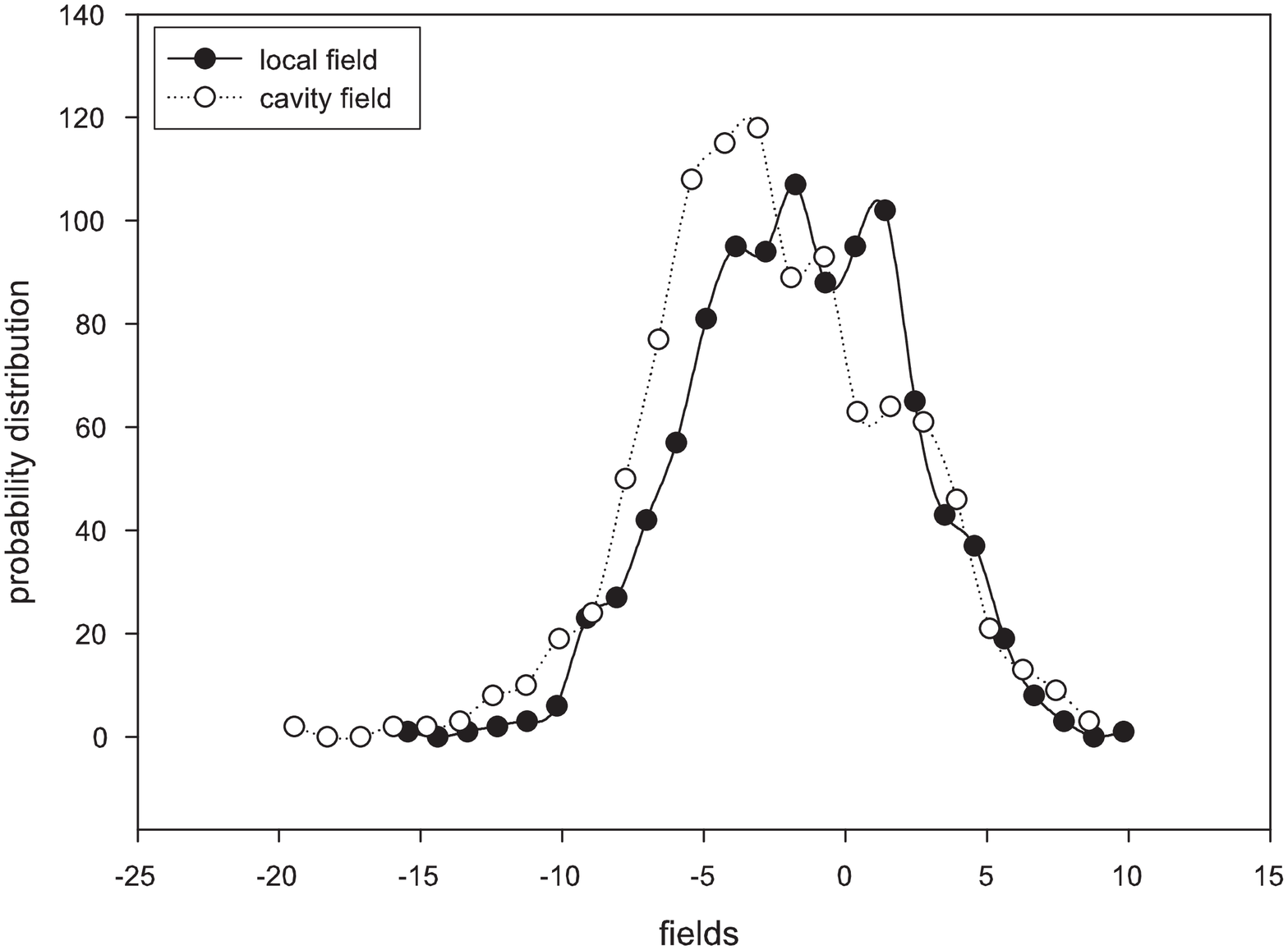}
 \caption{The local and the cavity fields in the presence of
noise, $\epsilon = 0.1$.}}
\end{figure}

This overlap is different for different species and depends on the
variance of the initial species distribution. This strange
behavior, found for some populations, is reminiscent of the
phase-transition phenomenon, and it is related to the following
peculiarities of our dynamical system: (i) all the populations are
positive; (ii) different initial conditions drive the ecosystem
into different attraction basins; and (iii)the complex structure
of the attraction basins of our dynamical system. While, in the
presence of noise, all the populations seem to be equivalent from
the dynamical point of view, some populations, in the absence of
external noise, have an asymptotical dynamical behavior such that
they significantly influence the dynamics of other species.

\section{Conclusions}
 We analyzed the nonlinear relaxation of an ecosystem
composed by N interacting species. We obtain the quasi-stationary
probability distribution of the population in the presence of
multiplicative noise. By using an approximation of the integral
equation, which gives the stochastic evolution of the system, we
obtained asymptotic behavior for different nonlinear relaxation
regimes. We observe an interesting phenomenon: the local and the
cavity fields, whose probability distributions are different in
the absence of noise, coincide for some populations in the
presence of multiplicative noise. This phenomenon can be ascribed
to the peculiarity of the dynamical system ($x_i > 0,$ always),
the influence of different initial conditions on the asymptotic
regime, and the complex structure of the attraction basins of our
ecosystem.

\section*{Acknowledgements}
This work was supported by MIUR, INFM and INTAS (project
2001-450).

%


\begin{thebibliography}{99}

\bibitem{Mur93}
    J. D. Murray, \emph{Mathematical Biology}, 2nd ed. (Springer, New York, 1993)
    \textbf{54} (1996), 706.

\bibitem{Kaw02}
    T. J. Kawecki and R. D. Holt, Am. \ Nat. \textbf{160} (2002), 333.

\bibitem{Gia01}
    I. Giardina, J. P. Bouchaud, M. Mezard, J. \ Phys. \ A: Math.
    Gen. \textbf{34} (2001), L245.

\bibitem{Esc04}
    C. Escudero, J. Buceta, F. J. de la Rubia, and Katja Lindenberg, Phys. \ Rev. \ E
    \textbf{69} (2004), 021908; H Rieger, J. \ Phys. \ A: Math.
    Gen. \textbf{22} (1989), 3447.

\bibitem{Droz04}
    Michel Droz and Andrzej P\c{e}kalski, Phys. \ Rev. \ E
    \textbf{69} (2004), 051912.

\bibitem{Soc01}
    J. E. S. Socolor, S. Richards, and W. G. Wilson, Phys. \ Rev. \ E
    \textbf{63} (2001), 041908.

\bibitem{Bar01}
    A. La Barbera and B. Spagnolo, Physica A \textbf{314} (2001), 120.

\bibitem{Val04}
    D. Valenti, A. Fiasconaro and B. Spagnolo, Acta \ Phys. \ Pol. B
    \textbf{35} (2004), 1481.

\bibitem{Fia04}
    A. Fiasconaro, D. Valenti and B. Spagnolo, Acta \ Phys. \ Pol. B
    \textbf{35} (2004), 1491.

\bibitem{Vil98}
    J. M. G. Vilar and R. V. Sol\'{e}, Phys. \ Rev. \ Lett.
    \textbf{80} (1998), 4099.

\bibitem{Spa02}
    B. Spagnolo and A. La Barbera, Physica A \textbf{315} (2002), 201.

\bibitem{Roz01}
    A. F. Rozenfeld Rozenfeld, C.J. Tessone, E. Albano, H.S. Wio, Phys. \ Lett. A
    \textbf{280} (2001), 45.

\bibitem{Spa03}
    B. Spagnolo A. Fiasconaro and D. Valenti, Fluct. \ Noise \ Lett. \textbf{3} (2003), L177.

\bibitem{Valb04}
    D. Valenti, A. Fiasconaro and B. Spagnolo, Physica A \textbf{331} (2004), 477.

\bibitem{Sci99}
    See the special section on \emph{"Complex Systems"}, Science \textbf{284} (1999),
    79-107; the special section on \emph{"Ecology through Time"}, Science \textbf{293} (2001)
    623-657.

\bibitem{Sci01}
    See the special section on \emph{"Ecology through Time"}, Science \textbf{293} (2001)
    623-657.

\bibitem{Spa04}
    B. Spagnolo D. Valenti, A. Fiasconaro, Math. \ Biosciences \ and Eng. \textbf{1} (2004), 185.

\bibitem{Ras00}
    D. F. Russel, L. AQ. Wilkens and F. Moss, Nature \textbf{402} (2000),291.

\bibitem{Ciu96}
    S. Ciuchi, F. de Pasquale and B. Spagnolo, Phys. \ Rev. \ E
    \textbf{54} (1996), 706; ibid. \textbf{47} (1993), 3915.

\bibitem{Spab02}
    B. Spagnolo, M. A. Cirone, A. La Barbera and F. de Pasquale, J.
    \ Phys.: Condens. Matter \textbf{14} (2002), 2247.

 \bibitem{Cir03}
    M. A. Cirone, F. de Pasquale and B. Spagnolo, Fractals \
    \textbf{11} (2003), 217.


\end{thebibliography}
\end{document}